\documentclass[aps,prb,twocolumn,showpacs,superscriptaddress]{revtex4}
\usepackage{graphicx}
\usepackage{amsfonts,amssymb,amsmath}
\usepackage{bm}
\begin{document}

\title{Nonlinear magnetoelastic behavior of the metastable bcc phases 
Co and Ni:
Importance of third-order contributions for bcc Ni}

\date{\today}
 
\author{Matej Komelj}
\email{matej.komelj@ijs.si}
\affiliation{Jo\v zef Stefan Institute, Jamova 39, SI-1000 Ljubljana,
  Slovenia}
\author{Manfred F\" ahnle}
\affiliation{Max-Planck-Institut f\" ur Metallforschung, Heisenbergstrasse 1,
D-70569 Stuttgart, Germany}
\begin{abstract}
The first- and second-order magnetoelastic coefficients of the  metastable
bcc phases Co and Ni are calculated  by using a combination of the 
phenomenological theory of nonlinear magnetoelasticity with the {\it ab-initio}
density functional electron theory. The magnetoelastic behavior of the bcc
phases is drastically different from that of the corresponding fcc phases.
The recently synthesized bcc phase of Ni appears to be an example of a material
for which third-order magnetoelastic effects are essential. 
\end{abstract}
\maketitle
In recent years it became possible to stabilize metastable phases of materials
by growth on appropriate substrates. For the transition metals Fe, Co and Ni
this is especially interesting because in these systems magnetism and structure
are closely related. Using molecular beam epitaxy, the fcc phases of Fe and 
Co could be stabilized on substrates at room temperature \cite{Pescia:1987}.
It became even possible to synthesize the bcc phase of Co \cite{Prinz:1985} on
various substrates (Ref. \onlinecite{Valvidares:2004} and references therein),
and most recently\cite{Tian:2005} the bcc phase of Ni. Both of these materials turned out to be
ferromagnetic at room temperature, with a magnetic moment per atom of
$1.53\>\mu_{\rm B}$ (Co) and $0.54\>\mu_{\rm B}$ (Ni). \par
From the viewpoint of technological applications of ultrathin magnetic films
the most important feature is the magnetic anisotropy. Because in general there
will be a lattice mismatch between the substrate and the magnetic film, 
magnetoelastic contributions to the magnetic anisotropy may be important. 
For instance, it has been suggested\cite{Subramanian:1995} that the in-plane 
anisotropy of bcc Co on GaAs is dominated by the magnetoelastic contribution,
although the epitaxial strains in this material are rather small, about 0.25\%.
For comparison, for bcc Co on Pt(001) the epitaxial strains are considerably
larger\cite{Valvidares:2004}  (-1.8\% in plane and 5.1\% out of plane). It is 
well known that for considerable epitaxial strains nonlinear  contributions
to the magnetoelastic energy become essential. This has been demonstrated 
experimentally by cantilever bending beam experiments (see, e.g., Refs. 
\onlinecite{Weber:1994,Sander:1999,Wedler:1999}): When changing the direction 
of the magnetization in the epitaxial film by changing the direction of the 
external magnetic field, the magnetostrictive stress $\Delta\sigma^{\rm m}$
along the cantilever axis changes, resulting in a detectable change of the 
bending of the film-substrate composite. In the framework of linear 
magnetoelastic theory, this change should be independent of the magnitude of 
the epitaxial strain and should be determined by the first-order magnetoelastic
coefficients, i.e., $B_1$ and $B_2$ for cubic materials. Experimentally, 
however, a linear dependence of $\Delta\sigma^{\rm m}$  on the strain was 
found, which was ascribed to nonlinear magnetoelastic effects. \par
For a proof of this conjecture a knowledge of the first- and second-order 
magnetoelastic coefficients of the respective bulk material is required.
The standard method to determine them is the ultrasonic pulse echo experiment.
Because the attainable strains in these experiments are very small, it is,
however, nearly impossible to explore the second-order magnetoelastic
coefficients by these experiments. The first confirmation of the conjecture
therefore was supplied by theory. By a combination of the phenomenological 
theory of nonlinear magnetoelasticity\cite{duTremolet:1993} with the
{\it ab-initio} density functional theory it has been shown
(see, for example, Refs. 
\onlinecite{Komelj:2002-1,Komelj:2002-2,Komelj:2002-3,Komelj:2001}
and references therein) that the second-order magnetoelastic contribution 
indeed may be very large, especially for the case of Fe. The theory was also 
able\cite{Komelj:2002-2} to suggest a complete set of six cantilever experiments
to determine the first-order ($B_1$ and $B_2$) and the second-order
($m_1^{\gamma,2}$, $m_2^{\gamma,2}$, $m_1^{\epsilon,2}$, $m_2^{\epsilon,2}$,
$m_3^{\gamma,2}$, $m_3^{\epsilon,2}$) magnetoelastic coefficients of a cubic
material. Thereby ($m_1^{\gamma,2}$, $m_2^{\gamma,2}$) is related to 
pure tensile strains, ($m_1^{\epsilon,2}$, $m_2^{\epsilon,2}$) to tensile and
shear strains, and ($m_3^{\gamma,2}$, $m_3^{\epsilon,2}$) to pure shear strains.
The first- and second-order coefficients have been 
calculated\cite{Komelj:2002-3} by the {\it ab-initio} electron theory for 
Fe, fcc Co, Ni, ${\rm Ni_3Fe}$ and CoFe. \par
The determination of the magnetoelastic coefficients is especially difficult
for metastable phases which can be synthesized only as epitaxial films on 
substrates, like fcc Co, bcc Co and bcc Ni, and in these cases the help of 
electron theory is very important. For the case of fcc Co, the theory 
has shown\cite{Komelj:2001} that the nonlinear magnetoelastic coupling 
coefficients are essential for the magnetostrictive strain but have only little
influence on the strain-induced out-of-plane anisotropy. In the present paper
we apply the theory to the case of bcc Co and bcc Ni. It will be shown that
in these systems the nonlinear magnetoelastic coefficients are again very large.
Furthermore, it will be shown that bcc Ni is the first example of a system for
which third-order magnetoelastic effects become relevant. \par
According to Ref. \onlinecite{Komelj:2002-2} the magnetoelastic coefficients 
may be obtained by exposing the cubic material to certain strain modes
$\epsilon_i$. Then the difference $\Delta e_i$ in the total energy 
per atom 
when changing the direction of the magnetization 
from $\alpha_i^1$ to $\alpha_i^2$ is calculated: \\
$i=1:\>\epsilon_1=\epsilon_{xx}=\epsilon_0,\>
\alpha_i^1=\langle100\rangle,\>\alpha_i^2=\langle 001\rangle$
\begin{equation}
\Delta e_1=B_1\epsilon_0+\left(B_1+{1\over 2} m_1^{\gamma,2}\right)\epsilon_0^2
\end{equation}
$i=2:\>\epsilon_2=\epsilon_{yy}=\epsilon_{zz}=\epsilon_0,\>
\alpha_i^1=\langle100\rangle,\>\alpha_i^2=\langle 001\rangle$
\begin{equation}
\Delta e_2=-B_1\epsilon_0+{1\over 2}\left(-B_1-m_1^{\gamma,2}+m_2^{\gamma,2}\right)
\epsilon_0^2
\end{equation}
$i=3:\>\epsilon_3=\epsilon_{xx}=\epsilon_{xy}=\epsilon_0,\>
\alpha_i^1=\langle010\rangle,\>\alpha_i^2=\langle110\rangle$
\begin{equation}
\Delta e_3=\left({B_1\over 2}+B_2\right)\epsilon_0+{1\over 2}\left(
{1\over 2}\left(B_1+m_1^{\gamma,2}\right)+B_2+m_2^{\epsilon,2}\right)
\epsilon_0^2
\end{equation}
$i=4:\>\epsilon_4=\epsilon_{zz}=\epsilon_{xy}=\epsilon_0,\>
\alpha_i^1=\langle010\rangle,\>\alpha_i^2=\langle110\rangle$
\begin{equation}
\Delta e_4=B_2\epsilon_0+{1\over 2}m_1^{\epsilon,2}
\epsilon_0^2
\end{equation}
$i=5:\>\epsilon_5=\epsilon_{xy}=\epsilon_0,\>
\alpha_i^1=\langle110\rangle,\>\alpha_i^2=\langle001\rangle$
\begin{equation}
\Delta e_5=-B_2\epsilon_0+{1\over 2}\left(m_3^{\gamma,2}-B_1\right)
\epsilon_0^2
\end{equation}
$i=6:\>\epsilon_6=\epsilon_{yz}=\epsilon_{zx}=\epsilon_0,\>
\alpha_i^1=\langle11\bar{2}\rangle,\>\alpha_i^2=\langle111\rangle$
\begin{equation}
\Delta e_6={8\over 3}B_2\epsilon_0+{1\over 12}\left(B_1+2B_2-m_3^{\gamma,2}+
2m_3^{\epsilon,2}\right)
\epsilon_0^2
\end{equation}
The coefficient $B_1$  and the pair ($m_1^{\gamma,2}$, $m_2^{\gamma,2}$)
of second-order coefficients are obtained from eqs.(1,2) by fitting parabola
to the data points for $\Delta e_1(\epsilon_0)$ and $\Delta e_2(\epsilon_0)$.
Similarly, the coefficients $B_2$, ($m_3^{\gamma,2}$, $m_3^{\epsilon,2}$)
are obtained from eqs.(5,6) by parabolic fits. Finally, the pair
($m_1^{\epsilon,2}$, $m_2^{\epsilon,2}$) is obtained from eqs.(3,4) via 
parabolic fits using the already determined coefficient  $m_1^{\gamma,2}$. 
As long as the parabolic fits represent the calculated data points 
$\Delta e_i(\epsilon_0)$ well, we can conclude that third-order magnetoelastic
effects can be neglected for the considered range of $\epsilon_0$. \par
The calculations of $\Delta e_i(\epsilon_0)$ were performed by applying the 
{\it ab-initio} density functional theory taking into account the spin-orbit 
coupling which is responsible for magnetoelasticity in a perturbative manner
using the second-variational method\cite{Singh:1994}. Furthermore, we use
the WIEN97 code\cite{Blaha:1990} which adopts the full-potential 
linearized-augmented-plane-wave method (FLAPW)\cite{Wimmer:1981} as well
as the local-spin-density approximation (LSDA)\cite{Perdew:1992-1} and
the generalized-gradient approximation (GGA)\cite{Perdew:1992-2} for the 
exchange-correlation functional. The strains $\epsilon_i$ were applied
with respect to the theoretically determined equilibrium lattice parameters 
$a=0.273\>(0.281)\>{\rm nm}$ for bcc Co and $a=0.273\>(0.279)\>{\rm nm}$ 
for bcc Ni in LSDA (GGA). The resulting LSDA (GGA) magnetic moments per
atom of $1.63\>(1.74)\>\mu_{\rm B}$ for bcc Co and of
$0.47\>(0.53)\>\mu_{\rm B}$ for bcc Ni are in agreement
with the experimental values of $1.53\>\mu_{\rm B}$ and $0.54\>\mu_{\rm B}$,
respectively.
\par
For the case of bcc Co all the data points $\Delta e_i(\epsilon_0)$ could be 
perfectly  fitted by parabola in the range $-0.03\le\epsilon_0\le 0.03$,
i.e., third-order effects can be neglected. Like for other materials
\cite{Komelj:2002-1,Komelj:2002-2,Komelj:2002-3,Komelj:2001} the discrepancy between 
LSDA and GGA may be
quite large. Because for the experimentally well investigated $B_1$ of bcc Co, 
fcc Ni and fcc Co the agreement with the GGA values was better than the 
agreement with LSDA, we concentrate in the following on the GGA results. 
For bcc Co the values of $B_1$ and $B_2$ are quite large as compared to bcc Fe,
fcc Ni and fcc Co. The second-order coefficients are also large. It is 
interesting that there
is a very large difference between bcc Co and fcc Co. This holds even 
for the first-order coefficients  $B_1$ and $B_2$ which magnitudes are 
considerably larger and of opposite sign for bcc Co as compared to fcc Co. 
In Ref. \onlinecite{Subramanian:1995} it has been assumed that for bcc Co the 
first-order magnetoelastic coefficients can be approximated by those of fcc Co,
in contrast to the results of our calculation. \par
\begin{figure*}
\includegraphics[scale=.75]{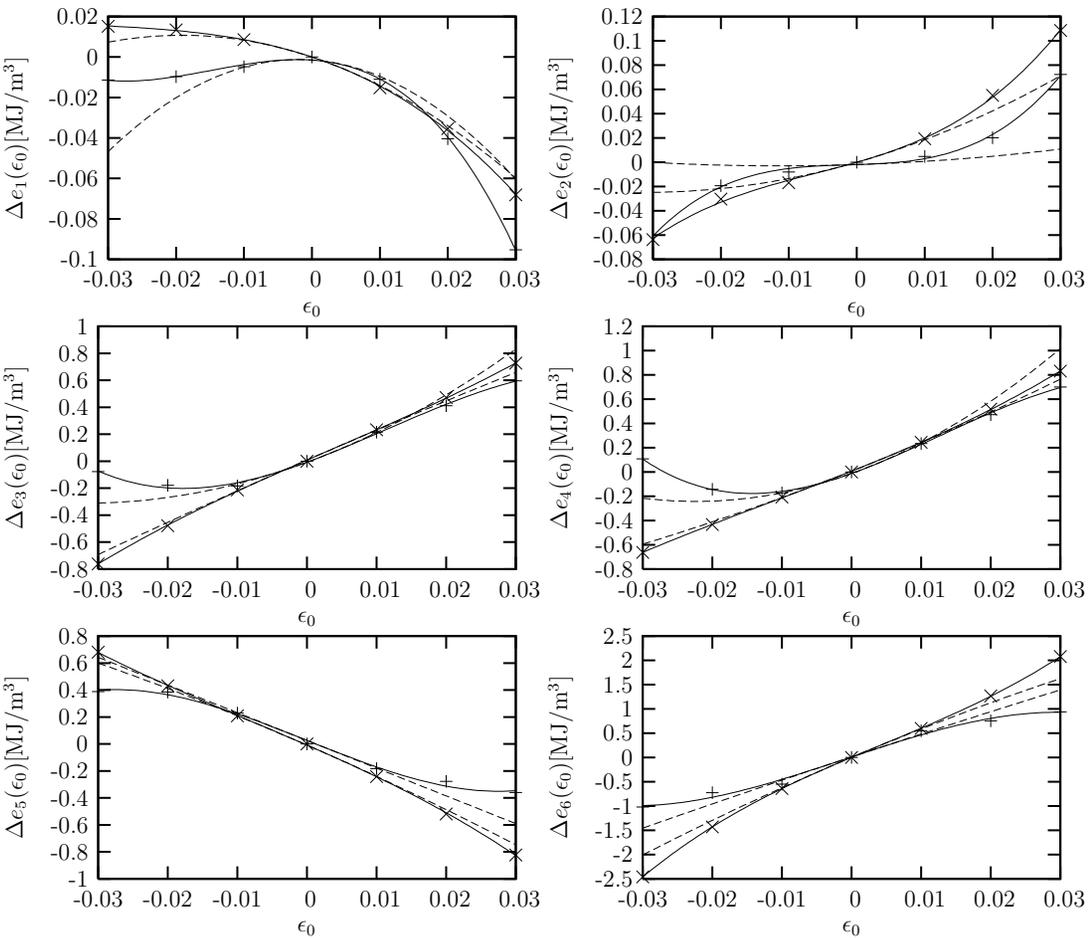}
\caption{The calculated functions $\Delta e_i(\epsilon_0)$ by applying LSDA
($+$) and GGA ($\times$). The solid lines are the third-order polynomials
fitted to the {\it ab-initio} calculated data points, whereas the dashed lines
represent the contribution up to the second order in $\epsilon_0$. }
\end{figure*}
The case of bcc Ni is even much more interesting because, as shown in Fig.
1, the data points for $\Delta e_i(\epsilon_0)$ show a drastic deviation
from a parabolic behavior in the range $-0.03\le\epsilon_0\le 0.03$. To the best
of our knowledge, bcc Ni therefore represents the first known material for which
third-order magnetoelastic effects become very important. Another surprising
result is that for bcc Ni the magnitude of $B_1$ is very small ($-1.3\>{\rm
MJ/m^3}$), much smaller than the one for fcc Ni ($10.2\>{\rm MJ/m^3}$). 
As in the case of Co, the magnetoelastic properties of the bcc phase are
drastically different from those of the fcc phase. This is in line with the 
experimental observations\cite{Tian:2005} that the cubic magnetic anisotropy
constant $K_1$  of bcc Ni is drastically different from the one of fcc Ni, and 
this was attributed to the different electronic band structures as found by
angle-resolved photoemission. \par
We hope that our prediction of strong third-order contributions to the 
magnetoelastic properties of bcc Ni will initiate an experimental investigation
by cantilever bending-beam experiments. To do this one has to grow epitaxial films
of bcc Ni with various average epitaxial strains $\epsilon_0$ which may be 
controlled with the film thickness \cite{Weber:1994} and then the change 
$\Delta\sigma^{\rm m}$ of the magnetostrictive stress due to a change of
the magnetization direction has to be measured. For the case that 
third-order  effects are relevant we expect a parabolic dependence:
\begin{equation}
\Delta\sigma^{\rm m}=a+D_1\epsilon_0+D_2\epsilon_0^2
\end{equation}
As discussed above, a linear dependence has been observed experimentally 
already for several materials. The observation of a quadratic contribution
would mean that for the first time a material was found for which the
third-order magnetoelastic contribution  is relevant. 
\begin{table*}
\begin{ruledtabular}
\begin{tabular}{llrrrrrrrrrrrrr}
&&$B_1$&$B_2$&$m_1^{\gamma,2}$&$m_2^{\gamma,2}$&$m_1^{\epsilon,2}$&
$m_2^{\epsilon,2}$&$m_3^{\gamma,2}$&$m_3^{\epsilon,2}$&$C_{11}$&$C_{12}$&
$C_{44}$&$\lambda_{100}$&$\lambda_{111}$\\
\hline
fcc Co\cite{Komelj:2002-3}&LSDA&-15.9&3.0&243&-53&81&102&759&796&
3.85&2.26&&6.7&\\
&GGA&-9.8&4.5&184&3&59&-41&862&1681&3.13&1.80&&4.9&\\
bcc Co&LSDA&61.5&-35.2&-672&575&357&-108&-363&336&2.1&2.95&1.78&48.2&6.6\\
&GGA&28.6&-39.6&-1013&973&148&51&-826&611&1.74&2.33&1.40&32.3&9.4\\
fcc Ni\cite{Komelj:2002-3}&LSDA&12.6&16.9&-117&23&168&-47&-2&388&3.63&2.20&&
-5.9&\\
&GGA&10.2&11.1&-95&71&90&-4&108&96&2.95&1.75&&-5.7&\\
bcc Ni&LSDA&-0.2&19.1&-116&-100&917&634&$\approx0$&-308&1.99&2.62&1.90&-0.2&-3.4\\
&GGA&-1.3&19.7&-28&-3&87&590&-75&-1347&1.52&2.32&1.63&-1.1&-4.0
\end{tabular}
\caption{The calculated magnetoelastic coefficients (in ${\rm MJ/m^3}$), 
elastic constants $C_{11}$, $C_{12}$ and $C_{44}$ (in $10^{11}\>{\rm N/m^2}$),
the magnetostrictive coefficients $\lambda_{100}=-2B_1/3(C_{11}-C_{12})$ and 
$\lambda_{111}=-B_2/3C_{44}$ (in units $10^{-5}$) Our calculated elastic 
constants for bcc Co and Ni agree nicely with those given in Ref. 
\onlinecite{Guo:2000}.}
\end{ruledtabular}
\end{table*}
\bibliography{text.bib}
\end{document}